\documentclass[twocolumn, 10pt, aps, floatfix, showpacs, prb, citeautoscript]{revtex4-1}
\usepackage{graphicx}
\usepackage{amsmath, amssymb}
\usepackage{bm}
\usepackage{xcolor}
\usepackage[colorlinks, citecolor={blue!50!black}, urlcolor={blue!50!black}, linkcolor={red!50!black}]{hyperref}
\usepackage{bookmark}
\usepackage{microtype}
\usepackage{units}

\newcommand{\comment}[2]{#2}

\begin{document}
\begin{abstract}
We report a new attractive critical point occurring in the Anderson localization scaling flow of symplectic models on fractals.
The scaling theory of Anderson localization predicts that in disordered symplectic two-dimensional systems  weak antilocalization effects lead to a metal-insulator transition. This transition is characterized by a repulsive critical point above which the system becomes metallic.
Fractals possess a non-integer scaling of conductance in the classical limit which can be continuously tuned by changing the fractal structure.
We demonstrate that in disordered symplectic Hamiltonians defined on fractals with classical conductance scaling $g \sim L^{-\varepsilon}$, for $0 < \varepsilon < \beta_\mathrm{max} \approx 0.15$, the metallic phase is replaced by a critical phase with a scale invariant conductance dependent on the fractal dimensionality.
Our results show that disordered fractals allow an explicit construction and verification of the $\varepsilon$ expansion.
\end{abstract}

\title{An attractive critical point from weak antilocalization on fractals}
\author{Doru Sticlet}
\author{Anton Akhmerov}
\affiliation{Kavli Institute of Nanoscience, Delft University of Technology, Lorentzweg 1, 2628 CJ Delft, The Netherlands}
\date{7 October 2015}

\maketitle

\emph{Introduction.}
The one-parameter scaling hypothesis~\cite{Abrahams1979} is central to the study of disordered electronic systems.
The hypothesis states that in disordered noninteracting systems the beta function $\beta = d \log g/d\log L$ determining the change of conductance $g$ with the system size $L$ is a universal function of $g$.
This hypothesis is known to be violated in topological insulators, where the topological invariant is the second scaling variable required to capture the scaling flow~\cite{Mong2012, Altland2015a}, and in systems where disorder itself is an irrelevant scaling variable~\cite{Altland2015, Syzranov2015, Gaerttner2015}.
Despite that, scaling flow of Anderson localization holds in an extremely broad range of systems~\cite{Lee1985, Evers2008}.

\begin{figure}[t]
\centering
\includegraphics[width=0.9\columnwidth]{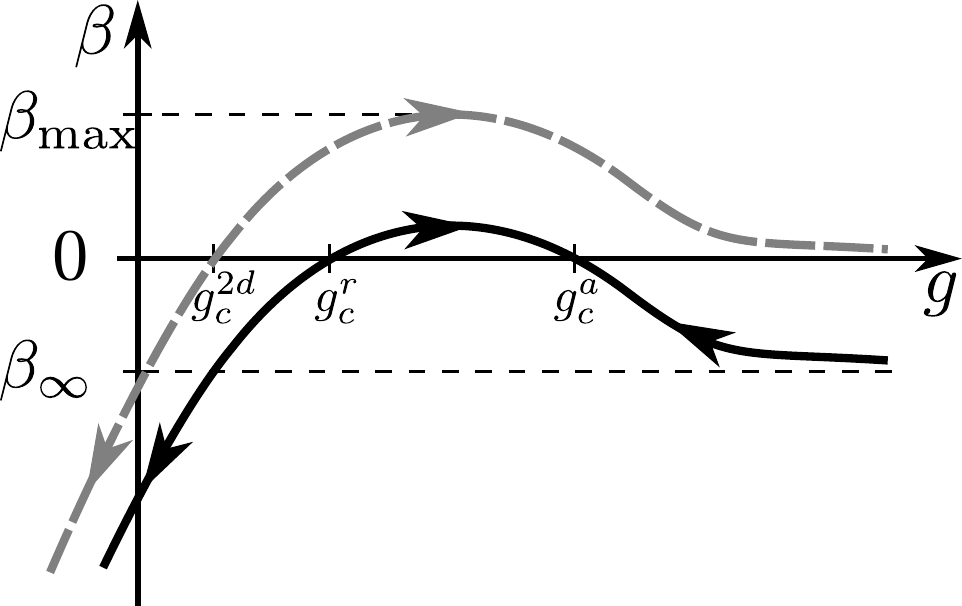}
\caption{The conductance flow ($\beta=d\log g/\log L$) as a function of dimensionless conductance $g$. There is a single repulsive critical point at $g = g_c^{2d}$ for symplectic systems in $2d$ (dashed gray line).
The conjectured flow for symplectic fractals with Hausdorff dimension lower than two (solid black line) can display a new attractive critical point $g_c^a$, in addition to the usual repulsive  point $g_c^r$.
The fractal is insulating in the diffusive limit, with a negative offset $\beta_\infty$ due to the fractal Ohm law.}
\label{fig:flow}
\end{figure}

The scaling flow has two universal regimes.
In the insulating regime $g \ll 1$ the exponential localization of the wave functions leads to a further decrease of conductance with the system size leading to $\beta\propto \log g + \mathrm{const.}$.
At high conductance, the beta function recovers the classical Ohm law, $\lim_{g\to\infty}\beta\equiv\beta_\infty=d-2$ with $d$ the Euclidean dimension.
A successful prediction of the theory was the occurrence of a metal-insulator transition in $3d$ as the flow passes between these two limits.
Later studies have refined the theory in the diffusive regime by taking into account quantum corrections to the Ohmic conductance~\cite{Lee1985,Evers2008}.
In time-reversal invariant systems with spin-orbit interactions, also called symplectic, the corrections to $g$ are positive, yielding weak antilocalization effects~\cite{Hikami1980}.
Consequently, these systems exhibit metal-insulator transition even in $2d$, with logarithmic corrections to conductance $g\propto\log L$, and a metallic phase at large conductance (Fig.~\ref{fig:flow}).

A successful approach in treating Anderson localizaton is $\varepsilon$ expansion, which treats $d$ as a continuous variable and constructs a series expansion of the scaling flow~\cite{Gorkov1979, Wegner1979, Abrahams1980, Wegner1980}.
The $\varepsilon$ expansion is a mathematical construct which is not expected to have a physical meaning when $d$ is not integer; nevertheless, fractals are examples of systems with non-integer dimensionality.
This motivates the main question of our work: Does the scaling hypothesis hold on fractals?
Scaling theory of phase transitions on fractals holds for Ising model~\cite{Gefen1980}, percolation transition~\cite{Havlin1983}, as well as metal-insulator transition of the Anderson model on bifractal lattices~\cite{Schreiber1996, Travenec2002}.
Tangential to our study are the investigations of shot noise in fractal resistor networks~\cite{Rammal1985,Groth2008} and quantum transport on clean Sierpi\'nski gaskets~\cite{Chakrabarti1996, Liu1999, Lin2002} and carpets~\cite{vanVeen2015}.

If the scaling hypothesis does hold, the simplest possible modification of the scaling flow on the fractal would be an overall shift of $\beta$ such that $\beta_\infty$ matches the properties of Ohm law on a fractal.
This leads us to the prediction: There exist fractals such that scaling flow of a symplectic model on them has a metal-insulator transition and an attractive critical point (shown in Fig.~\ref{fig:flow}).
To the best of our knowledge, appearance of a new type of a critical point on fractals is a unique property of Anderson scaling flow in symmetry classes allowing weak antilocalization, and it constitutes the main focus of our study.
Until now it has not been observed, despite numerical studies confirmed the presence of the usual metal-insulator transition on symplectic fractals with Hausdorff dimension lower than two.~\cite{Asada2006}

\emph{Analytical arguments for the presence of two critical points.}
Physical observables obey anomalous scaling laws in fractal systems, see Refs.~\onlinecite{Isichenko1992, Nakayama1994, Mandelbrot1982} for reviews.
Instead of the Euclidean dimension $d$ of the embedding space, these scaling laws are governed by the Hausdorff and the spectral dimension.
The Hausdorff dimension $d_h$ determines the scaling of the volume occupied by the fractal.
The spectral (fracton) dimension $d_s$ characterizes scaling of the low-energy density of states  of Laplace operator on the fractal $\rho(\omega)\propto \omega^{d_s/2-1}$~\cite{Alexander1982}, and therefore is relevant to diffusion and phonon dispersion on fractals.
While for Euclidean systems the exponents $d$, $d_h$, and $d_s$ are identical, in fractals they obey the inequality $d\ge d_h\ge d_s$~\cite{Rammal1983}.

The insulating limit of scaling flow is governed by the exponential localization of wave functions on almost decoupled orbitals, and applies whenever the growth of the number of $n$th nearest neighbors of an orbital is sub-exponential with $n$.
Therefore the scaling on fractals should hold in the insulating limit $g\ll 1$.
The classical limit $g\gg 1$ is governed by a diffusion equation.
Diffusion on fractals is slowed down so that the mean-square displacement of a random walker on fractals reads $\langle r^2\rangle\propto t^\gamma$, with the subdiffusion exponent $\gamma=d_s/d_h<1$~\cite{Gefen1981, Alexander1982}.
The Einstein relation links the diffusion and conductivity, and yields the scaling of diffusive conductance on fractals~\cite{Rammal1983, Ben-avraham1984, Nakayama1994}:
\begin{equation}\label{betainf}
g_0\propto L^{\beta_\infty},\quad \beta_\infty=d_h-2d_h/d_s.
\end{equation}
In particular, Ref.~\onlinecite{Rammal1983} makes an observation that the Eq.~\eqref{betainf} agrees with the scaling hypothesis of Anderson localization similar to the classical diffusive conductance on Euclidean lattices.

We compute the quantum correction to electronic transport on symplectic fractals using the return probability $p_0(t)$ after a time $t$~\cite{Chakravarty1986, Beenakker1991}.
In a fractal lattice the return probability is a function of $d_s$, $p_0(t)\propto t^{-d_s/2}$~\cite{Rammal1983, OShaughnessy1985}.
Therefore the conductance correction $\delta g$ to the fractal Ohm law reads:
\begin{equation}\label{dg0}
\frac{\delta g}{g_0}\propto\int_0^{\infty}dt\, p_0(t)e^{-t/\tau_\phi}\propto\tau_\phi^{1-d_s/2},
\end{equation}
with $\tau_\phi$ the phase-coherence time.
Calculating the prefactor of the integral is beyond the scope of our work.
It depends on fractal dimensions, but it is always positive for a Hamiltonian in the symplectic class.
Replacing $\tau_\phi$ with the typical time it takes a random walker to escape the system
$L^2 \propto \tau_\phi^{d_s/d_h}$, and using the definition of the scaling exponent Eq.~\eqref{betainf}, we conclude that the quantum correction to scaling is a scale-independent constant $g_c$, unlike a divergent correction $\delta g\propto \log L$ in $2d$.
Therefore the scaling function, calculated from the asymptotic form of conductance $g(L) = g_0 + g_c$, reads as
\begin{equation}\label{gcorr}
\beta=\beta_\infty-\beta_\infty g_c/g.
\end{equation}
This correction has the same perturbative dependence in $1/g$ as in the Euclidean lattices, and therefore its prefactor $\beta_\infty g_c \sim \mathcal{O}(1)$.

\comment{When $|\beta_\infty| \ll 1$, we expect to find an extra critical point.}
When $0 > \beta_\infty \gg -1$, the scaling function~\eqref{gcorr} vanishes at $g_c \sim -\beta_\infty^{-1} \gg 1$.
Since higher order quantum corrections to scaling $\delta g = \mathcal{O}(g^{-2})$ should be negligible when $g\gg 1$, the second critical point must indeed appear, and it has to be attractive, as shown in Fig.~\ref{fig:flow}.
In the following, numerical simulations will support this prediction.

\begin{figure}[t]
\includegraphics[width=\columnwidth]{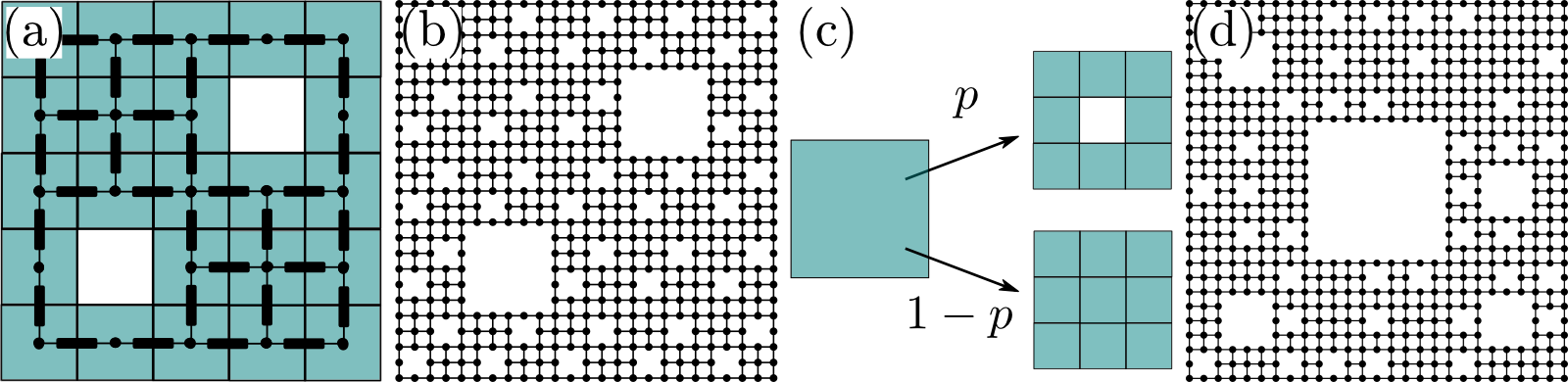}
\caption{Fractal patterns where we observe two critical points.
(a) Sierpi\'nski $5\times5$ pattern and (b) two-level lattice created using the pattern.
The fractal pattern is mapped to resistor networks (black rectangles denote resistors).
(c) Decimation procedure to create statistical fractals derived from $3\times 3$ Sierpi\'nski patterns and (d) a three-level lattice with $p=0.5$.
For statistical fractals, at any decimation step, there is a probability $p$ that a lattice block is carved with the fractal pattern.
The lattice turns from Euclidean ($p=0$) to regular fractal ($p=1$) by continuous variation of $p$.
}
\label{fig:resistor}
\end{figure}

\emph{Choice of the fractal.}
\comment{Limiting values of $\beta_\infty$ are bad, so we seek for a fractal with $\beta_\infty \approx -0.1$.}
The parameters of a fractal with two critical points are limited by two considerations.
On the one hand, if $|\beta_\infty|$ is too large, the quantum corrections to scaling become insufficiently strong to create a zero of the $\beta$ function.
Assuming that the main effect of changing the fractal exponents is an overall shift of the $\beta$ function provides the requirement $|\beta_\infty| < \beta_\textrm{max}$, with $\beta_\textrm{max}\approx 0.15$ the maximum~\cite{Asada2004} of the $\beta$ function in a symplectic $2d$ model.
On the other hand, $\beta_\infty$ must be negative for the second critical point to appear.
If $\beta_\infty$ approaches either of the limiting values, the scaling flow near the attractive critical point slows down, complicating the observation of criticality.
Additionally, when $\beta_\infty$ is close to zero, the attractive critical point occurs at large conductance $g$, requiring large system sizes, which are impractical for numerical simulations.
We therefore choose to use a fractal with $\beta_\infty \sim -0.1$.

In order to tune $\beta_\infty$, we consider two generalizations of Sierpi\'nski carpets.
In the first case, the fractal remains regular, with a number of subdivisions, larger than $3\times 3$ [see Fig.~\ref{fig:resistor}(a), (b)].
In the second case, statistical fractals are generated using a probabilistic subdivision rule, where the central subblock is removed with a probability $p$ (see Fig.~\ref{fig:resistor}(c), (d)).

Varying the amount and positions of removed cells in each subdivision or changing $p$ allows us to control $d_h$ and $d_s$.
In order to compute $\beta_\infty$, we construct finite-size versions of each fractal, replace the centers of each cell with nodes of a resistor network, and connect neighboring nodes with equal resistors, as shown in Fig.~\ref{fig:resistor}(a).
We set the potential of the leftmost modes to 0, and that of the rightmost to 1, and numerically solve the resulting Kirchhoff's equations for several system sizes and geometric disorder realizations in the case of statistical fractals.~\footnote{the source code and the resulting data are available as ancillary files to this manuscript.}
The Kirchoff's system of equations is defined and solved using Kwant package~\cite{Groth2014}. 
Finally, the length-independent $\beta_\infty$ follows by fitting the finite-size fractal results dependent on average $g(L)$ to
\begin{equation}\label{pow}
\beta_\textrm{classical}(L)\equiv\log_b
\frac{\langle g(L)\rangle}{\langle g(L/b)\rangle}
=\beta_\infty+cL^{-\mu},
\end{equation}
with $b$, the pattern magnification factor, and $c$ and $\mu$ constants.

This procedure yields results which agree with the exact value of $\beta_\infty$ for a Sierpi\'nski gasket~\footnote{see Supplemental Material}. For the classical $3\times 3$ Sierpi\'nski carpets, which lacks exact analytic results,\footnote{Ramification of a fractal measures the smallest number of bonds needed to be severed in order to detach any self-similar set from the rest of the network.
The Sierpi\'nski carpets have infinite ramifications since the connections of a self-similar set to the rest of the lattice grows with the size of its border.
If the ramification order is finite, it is possible to devise exact renormalization group techniques to determine the scaling of different physical observables.
This is the case for Sierpi\'nski gaskets, where block renormalization yields $\beta_\infty=\log_2(3/5)\approx -0.74$.}
we determine numerically $\beta_\infty\approx -0.2$, which lies within the bounds provided by the approximate renormalization group analysis~\cite{Gefen1984,Havlin1984,Wu1987}.
This result also means that classical Sierpi\'nski fractals will not host two critical points, which was a reason to search for alternative patterns. 
We find that the suitable value $\beta_\infty \approx -0.1$ is reached by the recursion pattern shown in Fig.~\ref{fig:resistor}(a) and statistical Sierpi\'nski carpets with $p=0.5$.
From resistor network simulations,\cite{Note1} we extract the classical conductance exponent in the $5\times 5$ pattern with $d_h=\log_5(23)$: $\beta_\infty\approx-0.1055$. In the $p=0.5$ statistical fractal with Hausdorff dimension $\bar d_h=\log_3(8.5)$, after averaging conductance for an ensemble of $10^4$ geometrically disordered lattices at each $L$, we find $\beta_\infty\approx-0.1063$.~\cite{Note2}

\emph{Quantum simulations of symplectic fractals.}
We investigate the scaling flow on fractals using the tight-binding Ando Hamiltonian~\cite{Evangelou1987, Ando1989}
\begin{equation}
H=\sum_{\bm{r}}\big[V_{\bm{r}}c^\dag_{\bm{r}}\sigma_0c_{\bm{r}}
-t\sum_{\bm{r}'\in\textrm{n.n.}} c^\dag_{\bm{r}'} e^{i\theta(\bm{r}' - \bm{r})\cdot\bm{\sigma}}c_{\bm{r}}\big],
\end{equation}
defined on sites $\bm{r}$ of a finite order fractal cut out of a square lattice with unit lattice constant, with $c^\dagger_{\bm{r}}$ and $c_{\bm{r}}$ the electron creation and annihilation operators.
Here $\sigma$ are the Pauli matrices acting in spin space, and $\theta=\pi/6$ is the spin-orbit coupling parameter.
The hoppings with amplitude $t=1$ are connecting the sites with their nearest neighbors on square lattice, if those are present in the fractal.
The random uncorrelated onsite potential $V_{\bm{r}}$ is uniformly distributed in the interval $[-V/2,V/2]$, with $V$ the disorder strength.
We attach leads to the leftmost and rightmost sites present in the system, and compute conductance using the Kwant package~\cite{Groth2014, Note1}.

\begin{figure}[t]
\centering
\includegraphics[width=\columnwidth]{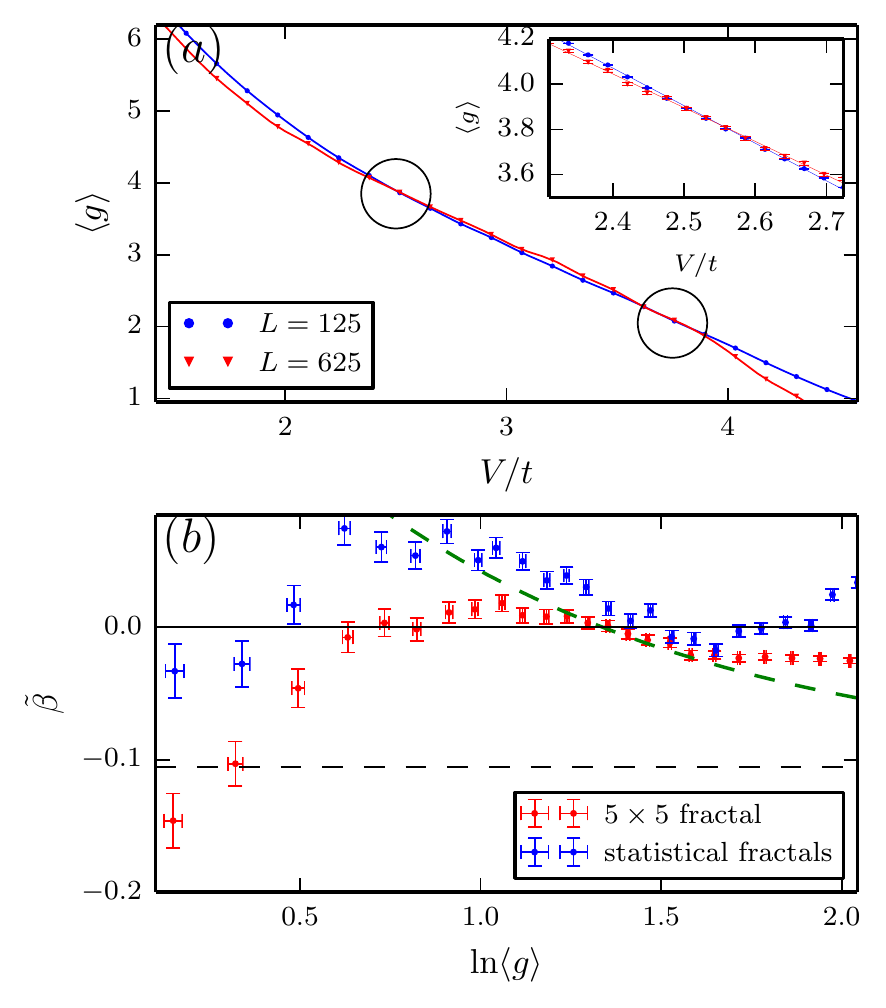}
\caption{(Color online) (a) Quantum transport results for a $5\times 5$ pattern Sierpi\'nski carpet.
Average dimensionless conductance $g$ as a function of disorder strength.
There are two critical points (encircled) and an intermediate metallic scaling regime.
There are $10^4$ disorder realizations for $L=125$ lattice, and $10^3$, for $L=625$.
The lines represent a cubic interpolation of experimental points (markers) to guide the eye.
The inset shows a zoom at $g_c^a$, where there are $10^5$ disorder realization for $L=125$ and $5\times 10^3$, for $L=625$.
The critical exponent is extracted from a linear fit of conductance curves near $g_c^a$ (inset lines).
The error bars represent the root-mean-square deviation.
(b) Approximate scaling function $\tilde\beta$ for $5\times 5$ fractals (red) and for the largest lattices sizes in statistical fractals $L=243, 729$ (blue).
The green dashed line is the prediction of Eqs.~\eqref{gcorr} with $g_c^a$ from $5\times 5$ pattern simulations.
The dashed black line denotes $\beta_\infty$.
}
\label{fig:5x5}
\end{figure}

We calculate the average conductance as a function of disorder strength and fractal size.
Our results for the fractal of Fig.~\ref{fig:resistor}(a) are shown in Fig.~\ref{fig:5x5}(a) for the sizes $L=125$, and $L=625$.
The corresponding data for the fractal of Fig.~\ref{fig:resistor}(c) with $p=0.5$ and fractal sizes 81, 243, and 729 is presented in the supplementary material and it exhibits the same trends.
We find that the conductance grows with size for intermediate disorder strengths, and drops both for high and low disorder.
The crossing of the curves at large disorder strength is the usual repulsive critical point marking the transition towards the strongly localized phase at $g<g_c^r$~\cite{Asada2006}.
The crossing of conductance curves at lower disorder strength realizes the attractive critical point $g_c^a$.
We fit the $g(L)$ in the vicinity of the attractive critical point with $g(L) = g_c^a+c (V-V_c)L^{1/\nu_a}$, with $V_c$, the critical disorder, $\nu_a$, the critical exponent, and $c$ a constant.
The resulting values are $1/\nu_a = -0.0668 \pm 0.00006$, $g^a_c= 3.811 \pm 0.002$ for the deterministic fractals, while for the statistical ones,~\cite{Note2} $1/\nu_a = -0.0847 \pm 0.0005$, and $g_c^a = 3.598\pm 0.026$ (for $L=243, 729$ lattices).
There are several sources of error which account for observed deviation of $\nu_a$ from the prediction given in Eq.~\eqref{gcorr} $1/\nu_a=\beta_\infty$.
Specifically, the $\mathcal O(g^{-2})$ corrections to the scaling function due to strong localization should make $|\nu_a|$ larger, while the finite-size corrections due to a finite mean free path may affect the obtained value of $\nu_a$ either way.

Since the numerics only limits the system sizes to two or three values, we calculate the approximate $\beta$ function $\tilde\beta=\log[g(L_2)/g(L_1)]/\log(L_2/L_1)$, with $L_2$ the largest available system size, and $L_1 = L_2 / b$.
This approximation is appropriate despite $b$ not being infinitesimally small since the conductance only changes slowly with length.
On the other hand, separating the finite-size corrections to the $\beta$ function requires more system sizes and larger computational resources.
The dependence of $\tilde\beta$ on $[g(L_2) + g(L_1)]/2$ is shown in Fig.~\ref{fig:5x5}(b) for both fractal types that we study.
We observe that both curves agree with the predicted qualitative scaling flow, and show a clear presence of two critical points.
Also in agreement with our expectations, the deviations from the predictions of weak antilocalization become significant not only at low conductances due to strong localization, but also at large conductance due to the increase of mean free path and finite-size effects.

\emph{Summary.}
We presented analytical arguments for validity of the scaling hypothesis on fractals, and showed that the competition between the positive quantum corrections to conductance and the diffusive conductance scaling can lead to the occurrence of an additional attractive critical point.
By tuning the fractal dimensions to the optimal regime with $\beta_\infty \approx -0.1$, we have also observed this critical point using numerical simulations.
Our findings are the first example of appearance of phases on fractals that cannot be observed in integer dimensions.

While the main relevance of our work is theoretical, the scaling of conductance can be observed experimentally using a patterned low effect mass quantum well with high spin-orbit coupling such as InAs or InSb at ultra-low temperatures.
With spin-orbit length $l_\textrm{SO}\approx \unit[100]{nm}$ at these structures, it becomes possible to aim for a \unit[30]{nm} feature size, while the dephasing length can be of an order of microns at lowest accessible temperatures.
A natural further question to investigate is how the multifractal properties of the wave functions at the critical point are tied to the fractal dimensions $d_h$ and $d_s$ of the parent fractal.
Our work can be straightforwardly adapted to other symmetry classes supporting weak antilocalization, such as the thermal metal phase characterized by the presence of particle-hole symmetry and absence of time reversal.
Finally, we expect that further analytical progress is possible using $d$-dimensional Sierpi\'nski gaskets due to their finite ramification.
When $d\to \infty$, the conductance scaling $\beta_\infty$ of Sierpi\'nski gaskets asymptotically approaches 0, offering a new way to analyze localization in regular $2d$ lattices.

\begin{acknowledgments}
We thank Yuval Oreg for drawing our attention to the topic, and Adriaan Vuik for code revison.
This research was supported by the Foundation for Fundamental Research on Matter (FOM), the Netherlands Organization for Scientific Research (NWO/OCW) as part of the Frontiers of Nanoscience program, and an ERC Starting Grant.
\end{acknowledgments}

\appendix
\section{Conductance scaling exponent from resistor networks}
\begin{figure*}[ht]
\centering
\includegraphics[width=\textwidth]{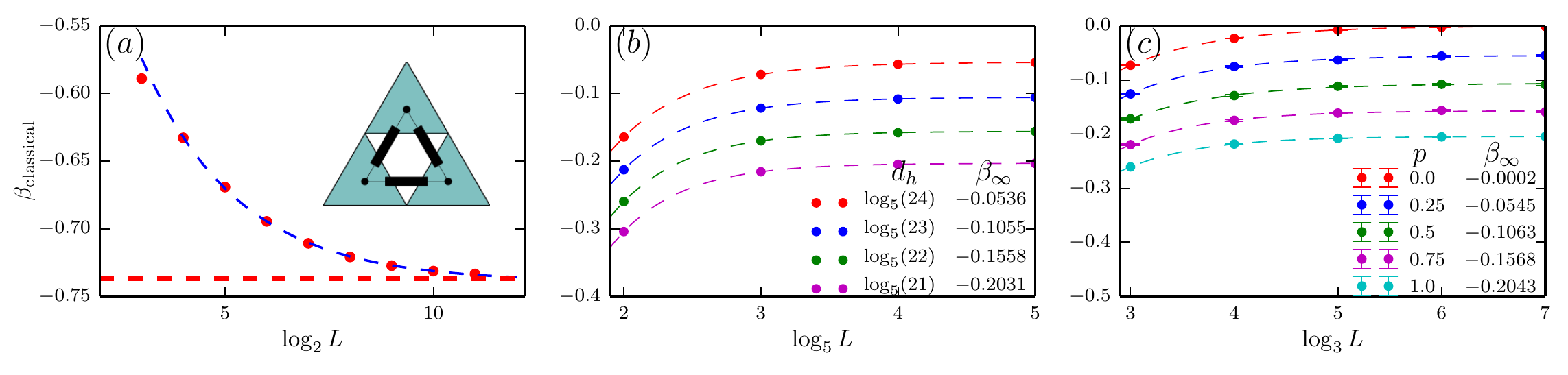}
\caption{(Color online) Finite-size effects of conductance scaling exponents extracted from resistor networks. The experimental points converge to the length-independent $\beta_\infty$ when increasing the lattice size $L$.
(a) For Sierpi\'nski gaskets with side length $L$, the conductance tends to the exact RG analytical value (dashed red line). The fitting experimental points (red) with Eq.~\eqref{powsm} yields accurate results with respect to the known analytical result (red dashed line) to third decimal (numerical: $\beta_\infty=-0.739\pm 0.0006$; exact analytic result: $\beta_\infty=\log_2(\frac{3}{5})\approx -0.73697$). The inset shows the fractal pattern.
(b) Conductance scaling on $5\times 5$ Sierpi\'nski carpets for different Hausdorff dimensions near $d_h=2$. The fractals explored in the body of the article have $d_h=\log_5(23)$ and $\beta_\infty>-0.15$.
(c) Conductance exponent for statistical fractal lattices built on Sierpi\'nski carpets. For different probabilities $p$, $\beta_\infty$ interpolates as expected between the Ohmic $2d$ result for a Euclidean lattice ($p=0$) and the regular Sierpi\'nski carpet ($p=1$).
For each $p\in(0,1)$ and $L$, conductance is averaged over $10^4$ randomly generated resistor networks. The error bars represent the root-mean-square deviation for the case of statistical fractals with $p\in (0,1)$.
The model studied in the article body has the exponent $\beta_\infty=-0.10626\pm 0.0016$, determined by fitting the experimental points with the power law~\eqref{powsm}.
}
\label{fig:ratios}
\end{figure*}

We use a decimation procedure to create the fractal lattice from a regular Euclidean lattice of edge size $L=ab^{n}$.
The lattice constant $a=1$, and it constitutes the cutoff length, while $b$ is the magnification factor of the fractal pattern. For example, the Sierpi\'ski gaskets have $b=2$ (Fig.~\ref{fig:ratios}(a)) and regular $3\times 3$ carpets, $b=3$.
The Euclidean lattice is divided into blocks, with some blocks removed according to the fractal pattern.
The procedure is then applied iteratively for each block until reaching the cutoff.
Therefore the fractal will present $n$ levels.

We map the lattice onto a resistor network by transforming each site into a node, and links into resistors.
The system of Kirchoff's equations and the Ohm's law reads:
\begin{eqnarray}
I_{ij}&=&G_{ij}(V_i-V_j),\\
\sum_j I_{ij} &=& I_\textrm{source},
\end{eqnarray}
with $V_i$ the potential of node $i$, $I_{ij}$ the current between nodes $i$ and $j$, and $G_{ij}=1/r$ the conductance between these nodes.
The current entering each node $I_\textrm{source}$ is nonzero only on the links of the system that are attached to the leads, where voltages are fixed.
We always consider a two lead situation with one lead at $V=0$ and the other at a finite voltage $V$.
The two terminal conductance $g$, computed for different fractals, is expressed through the full conductance matrix $G$, using the Schur complement formula:
\begin{equation}
\label{schur}
g=\sum_{ij}(G_{LL}-G_{LB}G_{BB}^{-1}G_{BL})_{ij},
\end{equation}
where indices $i, j$ run over the nodes connected to the left (source) lead.
The matrices $G_{LL}$, $G_{LB}$, and $G_{BB}$, are subblocks of the full conductance matrix $G$, with indices indicating whether they contain conductances among nodes connected to the source lead $(LL)$, between nodes connected to the left lead to internal (bulk) sites $(LB)$, or among the internal nodes $(BB)$.

Since only a few levels are available in numerical simulations, the length-independent conductance exponent $\beta_\infty$ follows by studying the convergence of numerical results for increasingly larger lattices. We find that there is a power law evolution towards $\beta_\infty$:
\begin{equation}\label{powsm}
\beta_\textrm{classical}(L)\equiv\log_b
\frac{\langle g(L)\rangle}{\langle g(L/b)\rangle}
=\beta_\infty+cL^{-\mu},\quad \mu>0.
\end{equation}
In the case of the Sierpi\'nski gasket, the analytical result for $\beta_\infty=\log_2 3/5$ agrees with the numerical solution (Fig.~\ref{fig:ratios}(a)).

Since, the $\beta_\infty$ of the Sierpi\'nski gasket is much smaller than the value required to see two critical points we consider first Sierpi\'nski $3\times 3$ carpets.
Since the number of external bonds of a self-similar set grows linearly with its size, it is not possible in this case to obtain the scaling exponent analytically. 
Nevertheless, bounds for the scaling exponent are determined using an approximate RG procedure.
Following Ref.~\onlinecite{Havlin1984}, we obtain $\beta_\infty\in(\log_3(2/3),\log_3(6/7))$. The upper bound is $\approx -0.14$ and it is close to desired values for $\beta_\infty$.
Mapping the system to a resistor network and solving the Kirchhoff's laws allows us to determine $\beta_\infty\approx -0.2$ (Fig.~\ref{fig:ratios}(c)).
This result indicates that the two critical points will not occur in this model.

Treating larger Sierpi\'nski patterns proceeds in a similar way to the $3\times 3$ case, and yields the conductance scaling exponent $\beta_\infty$ (Fig.~\ref{fig:ratios}(b)). Here we probe different patterns with Hausdorff dimensions close to 2, by varying the number of missing self-similar blocks in the fractal pattern. Results for conductance exponent are presented in Fig.~\ref{fig:ratios}(b). For quantum transport simulations, we focus on patterns with $d_h=\log_5(23)$. These comply with the lower bound of the scaling exponent $\beta_\infty>-0.15$, but $\beta_\infty$ is far enough from $d_h=2$, that the critical point is not pushed to very large conductance unattainable in numerical quantum transport simulations (the case of $d_h=\log_5(24)$).  

For statistical fractals, there is no approximate RG available. Consequently, we generate according to a stochastic process described an ensemble of $10^4$ disorder networks for a given length $L$ and probability $p$.
Fig.~\ref{fig:ratios}(c) shows the finite-size effects on the scaling exponent. For $p\ne 0$ or 1 the results interpolate between the Euclidean lattice and the geometrically clean Sierpi\'nski carpet. The figure also indicates that finite-size effects can be neglected for large lattices and the results converge to a length-independent exponent $\beta_\infty$ with the power law~\eqref{powsm}. Here we focus on statistical fractals with $p=0.5$ which have average Hausdorff dimension close to three decimals to the one studied for the $5\times 5$ pattern with $d_h=\log_5(23)$.

\section{Hausdorff dimension of statistically self-similar fractals}
\label{app:haus}
The Hausdorff dimension $d_h$ of a deterministic fractal embedded in $2d$ follows from the change in its area $A$ with a scale factor $b$ 
\begin{equation}\label{scal}
A(bL)=b^{d_h} A(L),
\end{equation}
where $L$ is the linear dimension. For a square fractal pattern $b\times b$ with $m$ missing blocks, the Hausdorff dimension reads from Eq.~\eqref{scal}:
\begin{equation}
d_h=\log_b(b^2-m).
\end{equation}

For statistical fractals, let us consider the average area and its scaling for an ensemble of $N$ elements, generated with our stochastic process for a given probability $p$.
The resulting average area after scaling with $b$ reads as
\begin{equation}
\langle A(bL)\rangle\approx \big[pNb^{d_h}\langle A(L)\rangle+(1-p)N b^2
\langle A(L)\rangle\big]/N,
\end{equation}
since there will be approximately $pN$ times when the fractal scaling holds, and $(1-p)N$ times when the Euclidean one holds. 
The above formula becomes an exact equality in the limit $N\to\infty$.
Therefore, from the scaling law for average areas: $\langle A(bL)\rangle=b^{\bar d_h}\langle A(L)\rangle$, it follows that the average Hausdorff dimension is
\begin{equation}
\bar d_h(p)=\log_b(b^2-pm).
\end{equation}
Such formula is readily generalizable to higher dimensions. For our particular case, the Siperpi\'nski carpet has a $3\times 3$ pattern and a missing central block ($b=3$, $m=1$). Therefore we recover the result in the body of the article.

\section{A statistical fractal with two critical points}

This section contains the visual representations of the data reported in the body of the article for the case of statistical fractals with $p=0.5$.

The Fig.~\ref{fig:ratios}(c) shows the case of resistor networks.
Scaling of the average conductance with $L$, $g(L)=3^{\beta_\infty}g(L/3)$ for an ensemble of $10^4$ geometrically disordered lattices determines the exponent $\beta_\infty$.
Fitting the experimental points with the power law~\eqref{powsm} gives the result $\beta_\infty=-0.10626\pm 0.0016$.

\begin{figure}[ht]
\includegraphics[width=\columnwidth]{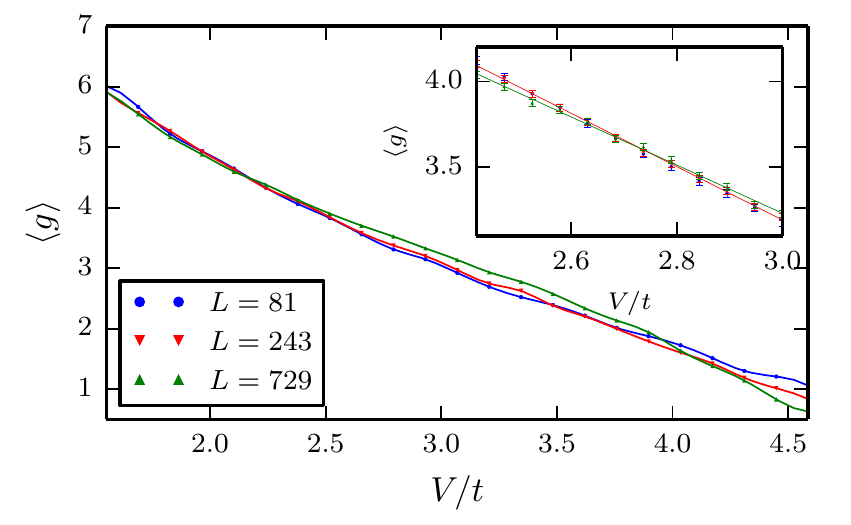}
\caption{(Color online)
Quantum transport simulations of the symplectic Ando model on statistical fractal networks with $p=0.5$.
Average conductance as a function of disorder strength for different lattice sizes $L$.
The solid lines are a cubic interpolation of the experimental points (markers).
The curves show two critical points where scaling changes. At high disorder there is the usual repulsive point $g_c^r$ marking the transition to the insulating phase. At lower disorder the new attractive critical point $g_c^a$ occurs. There are $10^5$ disorder realization for $L=81$, $10^4$, for $L=243$, and $10^3$ for $L=729$. 
The inset shows a zoom near $g_c^a$, and a linear fit (solid lines) for the largest lattice from which of the critical exponent is determined. 
At the zoom, there are $10^5$ disorder realizations for $L=81$ lattice, $10^4$ for $L=243$, and $5\times 10^3$ for $L=729$.
The error bars represent the root-mean-square deviation.
}
\label{fig:0p5}
\end{figure}

Fig.~\ref{fig:0p5} represents the average conductance as a function of disorder strength in the Ando model. The conductance curves reveal the presence of two critical points: the attractive one, $g_c^a$ at higher conductance, and the usual $g_c^r$ at lower conductance.
The inset contains a zoom near the attractive critical point $g_c^a$, where the conductance varies linearly with $L$. By fitting the experimental points, we have obtained the critical exponent $\nu_a$ and the an approximate value for $g_c^a\approx 3.598 \pm 0.026$ as reported in the body of the article.

The two critical points are observable for $p\in (0.45,0.55)$. More exact bounds require further investigations.


\bibliographystyle{apsrev4-1}
\bibliography{andersonfractal}
\end{document}